# All-Optical Generation and Detection of Coherent Acoustic Vibrations in Single Gallium Phosphide Nanoantennas Probed Near the Anapole Excitation

Hilario D. Boggiano[1], Nicolás A. Roqueiro[1], Haizhong Zhang[2], Leonid Krivitsky[2], Emiliano Cortés[3], Stefan A. Maier[4,5], Andrea V. Bragas*[1,6], Arseniy Kuznetsov[2], and Gustavo Grinblat*[1,6]

[1] Universidad de Buenos Aires, Facultad de Ciencias Exactas y Naturales, Departamento de Física. 1428 Buenos Aires, Argentina.
[2] Institute of Materials Research and Engineering (IMRE), Agency for Science, Technology and Research (A*STAR), 2 Fusionopolis Way, Innovis #08-03, Singapore, 138634, Republic of Singapore.
[3] Chair in Hybrid Nanosystems, Nanoinstitute Munich, Faculty of Physics, Ludwig-Maximilians-Universität München, 80539 München, Germany
[4] School of Physics and Astronomy, Monash University, Clayton, Victoria 3800, Australia
[5] Department of Physics, Imperial College London, London SW7 2AZ, U.K.
[6] CONICET - Universidad de Buenos Aires, Instituto de Física de Buenos Aires (IFIBA). 1428 Buenos Aires, Argentina.

* Email for G.G.: grinblat@df.uba.ar, email for A.V.B.: bragas@df.uba.ar.

**Abstract**

Nanostructured high-index dielectrics have shown great promise as low-loss photonic platforms for wavefront control and enhancing optical nonlinearities. However, their potential as optomechanical resonators has remained unexplored. In this work, we investigate the generation and detection of coherent acoustic phonons in individual crystalline gallium phosphide nanodisks on silica in a pump-probe configuration. By pumping the dielectric above its bandgap energy and probing over its transparent region, we observe the radial breathing mode of the disk with an oscillation frequency around 10 GHz. We analyze the performance of nanoantennas of various sizes in the 300-600 nm diameter range at fixed 125 nm height and find that the detection efficiency is maximum near the fundamental anapole state, in agreement with numerical simulations. By comparing with reference gold nanodisk and nanorod plasmonic resonators, we find that the dielectric nanoantennas display a modulation amplitude up to ~5 times larger. We further demonstrate the launching of acoustic waves through the underlaying substrate and the mechanical coupling between two nanostructures placed 3 μm apart, laying the basis for photonic-phononic signal processing using dielectric nanoantennas.

**Introduction**

Plasmonic nanoantennas have been widely used as efficient local transducers of far-field electromagnetic radiation into acoustic energy, with operating acoustic frequencies in the hypersound regime. The ability of these structures to control the electromagnetic field distribution at the nanoscale, together with their high-frequency acoustic dynamical response, has led to a vast number of studies and applications in sensing,[1-4] all-optical control of surface acoustic waves,[5-8] ultrafast vibrational microscopy,[9,10] and more.[11] The generation and detection of these coherent acoustic vibrations are usually carried out by ultrafast optical means. Recently, it was theoretically proposed that dielectric nanoantennas could outperform their metallic counterparts in a non-degenerate pump-probe configuration.[12]

In plasmonic resonators, absorption of light pulses excites the electron population, which relaxes by Landau damping and then by electron–electron scattering, followed by thermal equilibrium with the crystal lattice, eventually leading to mechanical oscillations of the nanostructure in its normal modes.[13, 14] A comparable effect can be generated in high-index nanostructured dielectrics, as long as the excitation is performed at energies above the bandgap.[12] Generally speaking, the heating-induced dilation of the lattice modifies the size and shape of the optomechanical resonator, giving rise to coherent mechanical oscillations and synchronic periodic fluctuations of the optical resonances, which can be detected with a pulsed probe beam.[15] However, in contrast to plasmonics, dielectrics can exhibit optical modes with negligible linear absorption below bandgap energies,[16] enabling higher probe intensities for optical readout. Also, their typically higher melting point as compared to metals allows for larger pump intensities without structural reshaping. Moreover, they support a much richer spectrum of electric and magnetic states as compared to plasmonic nanostructures,[17] expanding design possibilities and potential applications.

Nanoantennas made of high nonlinear susceptibility and high refractive index dielectric materials such as silicon (Si), germanium (Ge), gallium arsenide (GaAs), aluminum gallium arsenide (AlGaAs), and gallium phosphide (GaP) have been widely used in the past years to enhance nonlinear optical processes such as second and third harmonic generation,[18-22] four-wave mixing,[23, 24] and more,[25] of interest for all-optical data processing and communication.[26] The excitation of high-frequency coherent acoustic phonons in these structures would enable coherent modulation of their nonlinear optical properties in the gigahertz regime, relevant for current and future technologies.

In this work, we experimentally study optically resonant crystalline GaP nanodisks and demonstrate efficient excitation and detection of coherent acoustic mechanical oscillations, as theoretically anticipated.[12] We explore 125 nm thick disks with diameters in the 300-600 nm range in a dual color pump-probe scheme and observe that the strongest signal occurs when probing near the anapole state, where the field is highly confined, and the scattering cross section is minimum. We find that increasing structure size decreases the mechanical frequency of the radial breathing mode from ~11 to ~8 GHz, in agreement with numerical simulations. Furthermore, we demonstrate the launching of hypersound waves through the underlying substrate and their detection by a spatially separated second dielectric receiver resonator.

**Results and Discussions**

The sample consists of crystalline GaP nanodisks with diameters ranging from $D$ = 330 nm to $D$ = 580 nm and height $H$ = 125 nm (scheme depicted in Figure 1a), placed on top of a 10 μm thick $SiO_2$ layer supported by a 150 μm thick sapphire substrate (fabrication specifics are provided in the Methods section). Note that different from standard plasmonic metal structures,[27] no adhesion layer is needed in this case between the nanostructures and the substrate, which may help reducing losses, impacting significantly in the mechanical response of the disks. Representative scanning electron microscopy (SEM) images of nanostructures of different sizes are shown in Figure 1b. First, to model the linear optical response of the fabricated nanoantennas, we simulated the optical cross-sections under linearly polarized plane wave illumination at normal incidence (-$z$ direction in Figure 1a) in the 380-1200 nm wavelength range. The simulations were performed in the frequency domain using the finite element method (FEM) solver COMSOL Multiphysics (see Methods section for details on numerical calculations). In Figure 1c, a map of the scattering cross-section versus the wavelength of the incident light and the disk diameter reveals the multi-resonant nature of the nanoantennas. Among various radiative and non-radiative Mie modes, a pronounced scattering minimum can be distinguished and assigned to the first-order anapole excitation (AE). This optical state originates from the interaction between the electric and toroidal dipole modes,[28] as confirmed by a multipolar decomposition of the scattering cross section (see section S3 of the Supporting Information), where the featured minimum emerges from the far-field destructive interference between the out of phase electric and toroidal dipole moments. As can be seen from the computed absorption cross-section in Figure 1d, the nanostructures exhibit negligible linear absorption in the visible spectrum range (from ~450 nm wavelength) and multiple optical modes with large absorption cross-sections in the near-UV region.

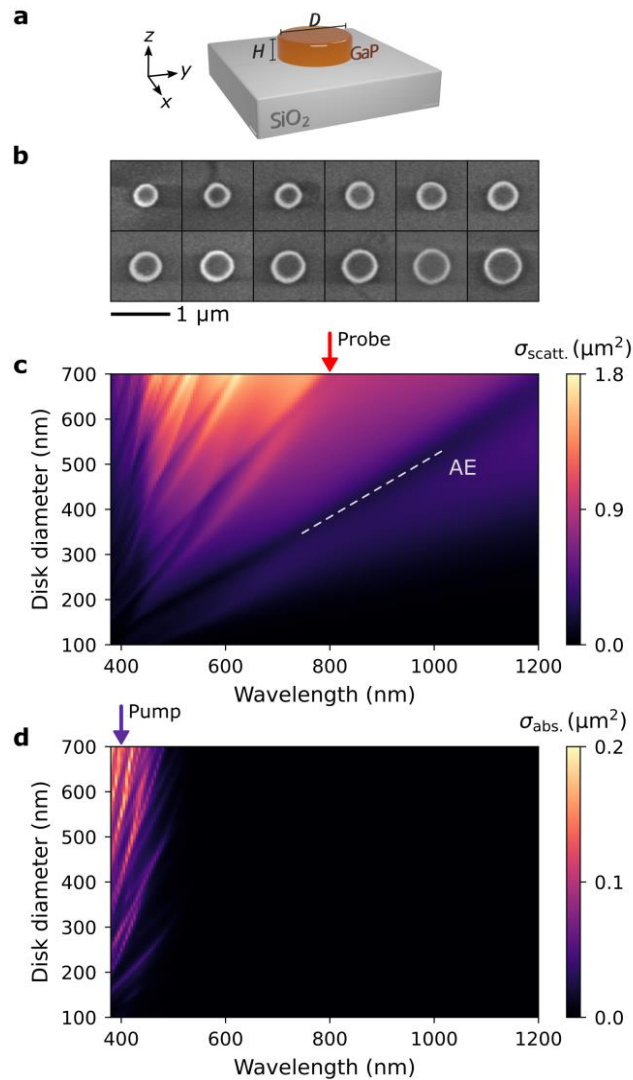

**Figure 1**. GaP nanostructures design and linear optical response. (a) Schematic view of the nanodisk antennas with different diameters $D$ and height $H$ = 125 nm on a SiO$_2$-on-sapphire substrate (10 μm and 150 μm thick, respectively). (b) Electron microscope representative images of the fabricated nanostructures with nominal diameters ranging from 360 nm (top-left) to 580 nm (bottom-right), by 20 nm steps. (c, d) Simulated scattering (c) and absorption (d) cross-sections as a function of incident wavelength and disk diameter. The dashed white line in panel (c) highlights the fundamental anapole excitation (AE), which manifests itself as a pronounced minimum in the scattering cross-section. Arrows at the top of each panel mark the pump and probe wavelengths.

To assess the ultrafast optomechanical response of the nanostructures, we performed two-color pump-probe measurements of individual particles in a transmission configuration. The second harmonic of a mode-locked Ti:sapphire laser with 95 MHz repetition rate and ~100 fs pulse duration was modulated and used to excite the dielectric above the bandgap energy, at ~400 nm wavelength. The fundamental laser output at ~800 nm wavelength was delayed and utilized to monitor changes in the scattering near the fundamental anapole state. Arrows at the top of Figure 1c,d indicate the pump and probe wavelengths, where the correlation between the excitation/detection wavelengths and the absorption/scattering of the nanostructures can be seen. Further information on the ultrafast measurements can be found in the Methods section.

Figure 2a exhibits the measured extinction (without pump beam) as a function of the disk diameter (circles) at fixed probe wavelength, showing very good agreement with the simulated scattering efficiency (solid line). The latter was computed as the linear scattering cross-section divided by the disk geometric cross-section. The experimental values correspond to $1-(T-T_B)$, $T$ being the measured transmission of the laser (centered at 795 nm wavelength) focused on a single nanoantenna using a 40× (0.6 numerical aperture) microscope objective. The transmission data was corrected by subtracting the transmission background $T_B$ outside the geometric cross-sectional area of the antenna, proportional to the laser power fraction illuminating the substrate surrounding the nanodisk. A pronounced dip in the extinction is observed around $D = 380$ nm and attributed to the strong suppression of the linear scattering at the anapole condition. This scattering minimum is accompanied by an electric field enhancement inside the nanostructure, as shown in Figure 2b. The inset exhibits the distinctive electric field distribution associated to the first-order AE.[28]

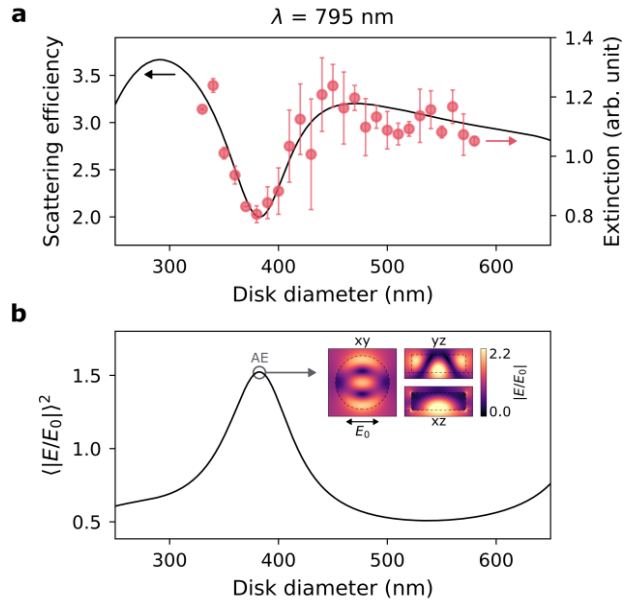

**Figure 2**. Experimental transmission and simulated electric field enhancement. (a) Simulated scattering efficiency and measured extinction (circles) of individual nanoantennas of different diameters at 795 nm wavelength illumination. The error bars correspond to the standard deviation of each measured set of nanodisks with the same nominal diameter. (b) Average electric field enhancement within the antenna volume. The inset shows the computed normalized electric field distribution for a $D = 382$ nm nanodisk at the AE ($\lambda = 795$ nm) under linearly polarized illumination in the direction indicated with an arrow ($E_0$). The *xy* field map was computed at half of the disk height, while the *yz* and *xz* distributions correspond to half-diameter cross-sections.

Figure 3a,b (top) shows the differential transmission ($\Delta T/T$) pump-probe measurements of two nanodisks of selected diameters, $D = 420$ nm and 360 nm, above and below the anapole state

condition at the detection wavelength, respectively (negative and positive slope wings of the AE in Figure 2a). In both cases, the signal background is well described by a biexponential decay, where the fastest positive component ($\tau_1 \sim 16$ ps) can be associated to the relaxation of the photogenerated free carriers,[29] while the long-decay component (which exceeds the time window studied) corresponds to the thermalization of the system. We also note that the modulation sign correlates with the slope sign of the scattering/extinction curve in Figure 2a. The ultrafast optical excitation and relaxation of photogenerated free carriers is followed by a rapid heating and consequent expansion of the crystal lattice that sets the nanoparticle in motion at its normal modes. In Figure 3c,d, the fast Fourier transform of the signal's residual (Figure 3a,b, bottom panels) reveals in both cases an oscillatory component at around 10 GHz, associated with the main radial mechanical mode of the nanostructures. By applying a high-pass Fourier filter with a cutoff frequency of 5 GHz, the phononic components can be distinguished from the low-frequency background (Figure 3e,f).

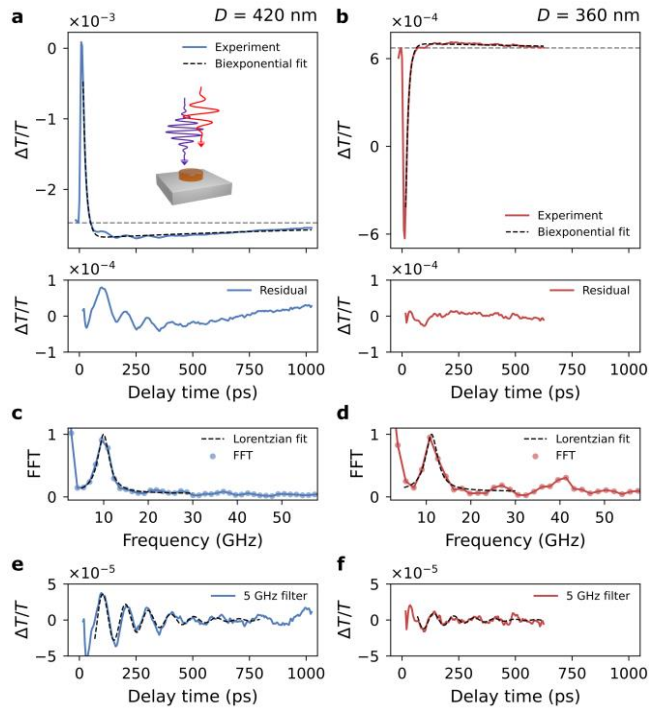

**Figure 3**. Pump-probe measurements on single GaP nanodisks. (a, b) Top: experimental differential probe transmission signals ($\Delta T/T$) of disks with diameter $D$ = 420 nm (a) and $D$ = 360 nm (b). Nanostructures are excited at ~400 nm wavelength (60 µW average power) and probed by a delayed ~800 nm pulse (15 µW average power). The black dashed line is a biexponential fit. The horizontal dashed line indicates the signal offset. The inset is an illustration of the pump-probe experiment on a single nanodisk. Bottom: residual signal after background subtraction, displaying oscillations. (c, d) Normalized fast Fourier transform (FFT) of the phononic signal in (a, b), along with a Lorentzian peak curve fit. (e, f) Same signal as at the bottom of (a, b) after applying a 5 GHz high-pass Fourier digital filter along with a single damped sinusoid fitting curve (dashed line).

Next, we show the simulated average radial displacement, $\langle u_r \rangle$, as a function of mechanical frequency and disk size, together with the measured normal frequencies, as seen in Figure 4a. The simulation displays two strong branches, where only the top one is experimentally observed, in very good agreement with the calculation. As the disk diameter increases, the mechanical frequency decreases from ~11 to ~8 GHz. The average Q-factor obtained for all nanodisk sizes, calculated as $Q = \pi \tau f$, where $\tau$ is the decay constant of the oscillatory signal of frequency $f$, is $Q = 6$ . Figure 4b exhibits instants of maximum deformation (expansion and contraction) of the radial breathing mode described by the top branch in Figure 4a. The lower branch originates from the coupling of the nanostructure to the substrate (see section S5 of the Supporting Information). As a fixed surface condition is assumed for the disk-substrate interface in the model (i.e., ideal coupling), a lower detection sensitivity is expected in the experiment. This, and an eventually weaker phonon-optical coupling for this mode, could be the reasons for its absence in the measurements.[30, 31]

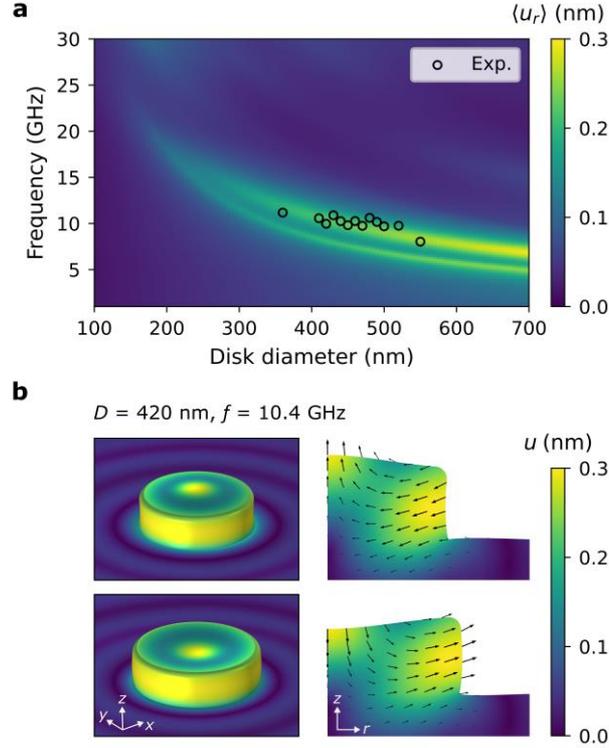

**Figure 4**. Elastic mechanical response. (a) Frequency-domain simulation of the average radial displacement, $\langle u_r \rangle$, for different disk diameters. A different representation of the same data is included in Section S4 of the Supporting Information. The experimental frequencies are shown as black empty circles. The frequency error (not displayed in the figure) obtained from curve fitting is in the order of ≤ 0.2 GHz. (b) 3D view (left) and cross-section (right) of the displacement field at the main acoustic resonance. A scale factor of 50× was applied to enlarge the deformation for visual clarity. An increase in the lattice temperature $\Delta T_L$ = 100 K was considered as the displacive excitation mechanism, and an isotropic loss factor $\eta_s$ = 0.1 was implemented to account for the internal damping.

Following the approach outlined by Yan et al.,[12] we compute a figure of merit $\eta$ to quantify the optomechanical excitation and detection efficiencies as follows:

$$\eta = \sigma_{\text{abs.}} |\partial_\lambda \sigma_{\text{scatt.}}| \langle u_r \rangle$$

where $\sigma_{\text{abs.}}$ is the absorption cross-section at the excitation wavelength of 400 nm (Figure 5a), $|\partial_\lambda \sigma_{\text{scatt.}}|$ is the absolute value of the derivative of the scattering cross-section $\sigma_{\text{scatt.}}$ with respect to the wavelength $\lambda$ at the detection wavelength of 800 nm (Figure 5b), and $\langle u_r \rangle$ is the average displacement in the radial direction for a given increase in the lattice temperature (Figure 5c). In Figure 5d, we compare the resulting quantity with the maximum absolute variation in transmission of the oscillatory signal, $\Delta T$, normalized by the pump power, $P_{\text{pump}}$, showing good agreement. The highest efficiency occurs for a diameter of 410 nm at which both $\sigma_{\text{abs.}}$ and $|\partial_\lambda \sigma_{\text{scatt.}}|$ display a peak. For reference, we include the results from resonant gold nanodisk and nanorod plasmonic antennas (horizontal lines), as taken from our previous reports,[3, 6] which show a modulation amplitude up to ~5 times smaller than that of the dielectric nanodisk, in consistency with recent predictions.[12]

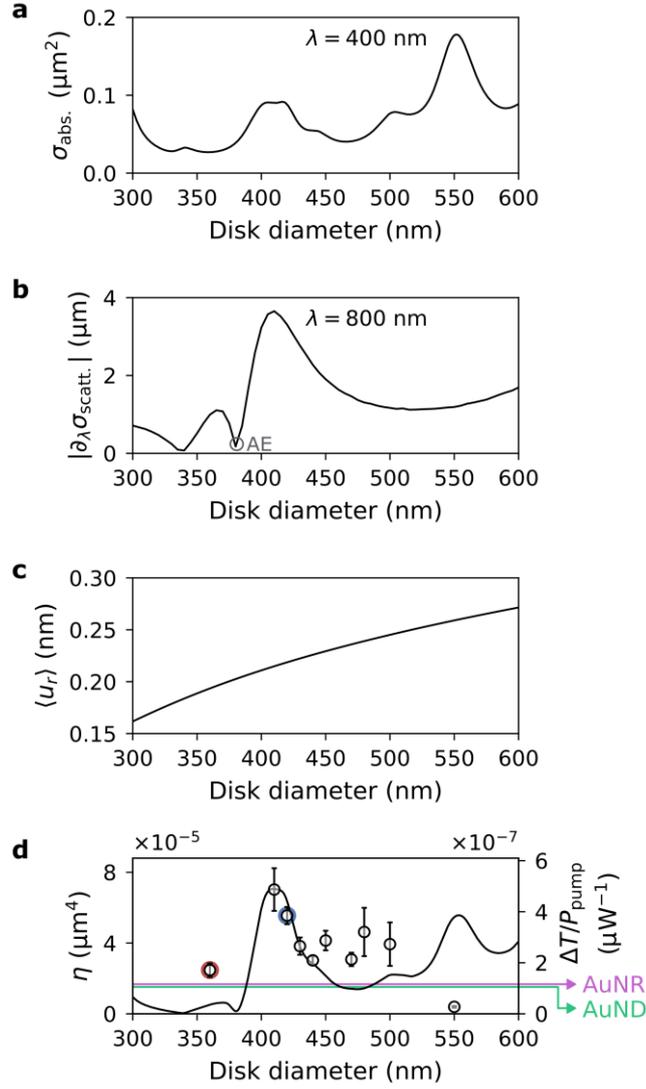

**Figure 5**. Figure of merit for the excitation and detection of coherent acoustic phonons. (a) Absorption cross-section $\sigma_{\text{abs.}}$ at the excitation wavelength ($\lambda = 400$ nm). (b) Derivative of the scattering cross-section $\sigma_{\text{scatt.}}$ with respect to wavelength evaluated at the detection wavelength ($\lambda = 800$ nm). (c) Average disk displacement in the radial direction $\langle u_r \rangle$ for a lattice temperature increase $\Delta T_{\text{L}} = 100$ K. (d) Figure of merit computed as $\eta = \sigma_{\text{abs.}} |\partial_\lambda \sigma_{\text{scatt.}}| \langle u_r \rangle$, along with the measured maximum amplitude of the transmission variation $\Delta T$, normalized by the pump power, $P_{\text{pump}}$. The highlighted red and blue diameters correspond to those in Figure 2. The horizontal lines correspond to measurements from resonant gold nanorod (AuNR) and nanodisk (AuND) antennas as taken from our previous works.[3, 6] The dimensions are 144 (length) × 67 (width) × 35 (height) nm for the nanorod and 140 nm diameter with 30 nm height for the nanodisk.

Finally, we also demonstrate the launching of acoustic waves through the underlying substrate as excited by the pumped GaP nanodisk, and their detection by probing over a second nanoantenna placed 3 µm away (see section S6 of the Supporting Information). Based on the arrival time of the mechanical wave from the emitter to the receiver, a propagation velocity of (6.0 ± 0.7) nm/ps is attained, which matches well with the speed of surface-skimming longitudinal (SSL) bulk phonons.[5, 32] This finding shows the potential for photonic-phononic signal processing using dielectric nanostructures.

**Conclusions**

In summary, we have shown that the optomechanical modulation of the anapole condition enables optically probing coherent acoustic vibrations in single GaP disk nanoantennas with high efficiency. The mechanical coupling between two ~200 nm radius nanoantennas with a center-to-center separation of 3 μm through the underlying substrate has also been described. The results of this investigation pave the way for the engineering of non-plasmonic nanostructures for the efficient generation, detection, and manipulation of ultrahigh frequency acoustic waves at the nanoscale. We anticipate that other dielectric materials, such as AlGaAs, could exhibit a higher performance due to its larger absorption coefficient in the near-UV region −while maintaining transparency in the near-infrared− and higher refractive index. Exploring higher quality factor optical resonances, such as higher order anapoles, Fano resonances, or quasi-bound states in the continuum could further improve the modulation amplitude. The next efforts should also be aimed at reducing external losses, in order to achieve higher mechanical quality factors, promoting the application of these nanoresonators.

**Methods**

Sample Fabrication
The GaP active layer (~400 nm) is first grown on a gallium arsenide (GaAs) substrate with an AlGaInP buffer layer by metal-organic chemical vapor deposition (MOCVD) to reduce lattice mismatch between the GaAs substrate and the GaP layer. This structure is then directly bonded to a sapphire substrate after depositing a $SiO_2$ layer on top of both surfaces. The AlGaInP/GaAs substrate is then removed by wet etching and the thickness of the GaP layer is thinned down to 125 nm. The fabrication of the GaP nanostructures with diameters ranging from $D$ = 330 nm to $D$ = 580 nm starts with a standard wafer cleaning procedure (acetone, iso-propyl alcohol and deionized water in that sequence under sonication), followed by $O_2$ and hexamethyldisilizane (HMDS) priming in order to increase the adhesion between GaP and the subsequent spin-coated electron-beam lithography (EBL) resist of hydrogen silsesquioxane (HSQ). After spin-coating of an HSQ layer with a thickness of~540 nm, electron beam lithography and development in 25% tetra-methyl ammonium hydroxide (TMAH) are carried out to define the nanostructures in HSQ resist. Inductively-coupled plasma reactive ion etching (ICP-RIE) with $N_2$ and $Cl_2$ is then used to transfer the HSQ patterns to the GaP layer. Finally, HSQ is removed to generate the designed GaP disk nanoantennas.

Numerical Simulations
Numerical calculations were performed using the Wave Optics and Structural Mechanics modules of the FEM solver software COMSOL Multiphysics. For linear optical simulations a spherical domain with perfectly matched layer (PML) boundary condition was implemented. The disk, atop the silica substrate, was located at the center and surrounded by air. The scattering field was solved using the analytical expressions of the reflected and transmitted background electric fields and then used to compute the optical cross-sections and electric field intensity enhancement by standard procedures. The linear elastic response of the nanostructures was obtained by solving the Navier's equation in the frequency domain. A thermal strain $\varepsilon_{\text{th}}$, proportional to the increase in the lattice temperature, $\Delta T_L$ = 100 K, was considered as the displacive excitation mechanism: $\varepsilon_{\text{th}} = \alpha \Delta T_L$, where $\alpha$ is the coefficient of linear thermal expansion. To account for the intrinsic damping mechanisms, an isotropic loss factor $\eta_s$ = 0.1 was implemented in the GaP domain. PMLs were used to truncate the computational domain, simulating an infinite substrate. The continuity of the stress and displacement fields was considered at all boundaries. Reference values for the properties of materials used in numerical calculations are given in sections S1 and S2 of the Supporting Information.

Pump-probe Experiments and Transmission Measurements

Transient transmission measurements were carried out using an ultrafast two-color pump-probe setup with a mode-locked Ti:Sapphire laser (KMLabs). The laser produces ∼100 fs pulses with a repetition rate of 95 MHz and an average output power of 300 mW, centered at 800 nm wavelength. The output beam is focused on a BBO crystal, and its second harmonic at 400 nm is used as the pump beam. The residual fundamental light is delayed with a motorized translation stage and used as the probe beam. The measurements were performed using lock-in detection by modulating the pump beam at 100 kHz frequency using an acousto-optic modulator. Both beams were focused onto the sample through the same objective (40×, NA = 0.6) with a spot radius ($e^{-2}$) of 1.1 μm (pump) and 0.8 μm (probe), with the help of a home-made dark-field microscope. The average pump and probe powers at the sample were 60 μW and 15 μW, respectively.

For the extinction characterization, the fundamental laser output, centered at 795 nm wavelength, was focused onto individual nanostructures. The transmitted light was collected by a 10× (NA = 0.28) microscope objective and directed to a photoreceiver.

**Supporting Information**

Refractive index and extinction coefficient for crystalline GaP; Reference values for properties of materials used in numerical simulations; Multipolar decomposition; Frequency and displacement at the main mechanical resonance as a function of the disk diameter; Frequency-domain simulation of the average displacement in the z-direction; Propagation of acoustic waves.

**Acknowledgments**

This work was partially funded by PICT 2021 IA 363 and PICT 2021 GRF TI 349 (ANPCyT), PIP 112 20200101465 (CONICET), UBACyT Proyecto 20020220200078BA. A.V.B acknowledges funding support from the Alexander von Humboldt Foundation through the Georg Forster Award. We acknowledge funding and support from the Deutsche Forschungsgemeinschaft (DFG) under Germany´s Excellence Strategy – EXC 2089/1 – 390776260, the Bavarian program Solar Technologies Go Hybrid (SolTech) and the Center for NanoScience (CeNS) based at LMU Munich. S.A.M. acknowledges the Lee-Lucas Chair in Physics. H.D.B. acknowledges Ornella Colmegna for assistance with multipole expansion calculations.

**Conflict of Interests**

The authors declare no competing financial interest.

**Supporting Information for:**

**All-Optical Generation and Detection of Coherent Acoustic Vibrations in Single Gallium Phosphide Nanoantennas Probed Near the Anapole Excitation**

Hilario D. Boggiano[1], Nicolás A. Roqueiro[1], Haizhong Zhang[2], Leonid Krivitsky[2], Emiliano Cortés[3], Stefan A. Maier[4,5], Andrea V. Bragas*[1,6], Arseniy Kuznetsov[2], and Gustavo Grinblat*[1,6]

[1] Universidad de Buenos Aires, Facultad de Ciencias Exactas y Naturales, Departamento de Física. 1428 Buenos Aires, Argentina.
[2] Institute of Materials Research and Engineering (IMRE), Agency for Science, Technology and Research (A*STAR), 2 Fusionopolis Way, Innovis #08-03, Singapore, 138634, Republic of Singapore.
[3] Chair in Hybrid Nanosystems, Nanoinstitute Munich, Faculty of Physics, Ludwig-Maximilians-Universität München, 80539 München, Germany
[4] School of Physics and Astronomy, Monash University, Clayton, Victoria 3800, Australia
[5] Department of Physics, Imperial College London, London SW7 2AZ, U.K.
[6] CONICET - Universidad de Buenos Aires, Instituto de Física de Buenos Aires (IFIBA). 1428 Buenos Aires, Argentina.

* Email for G.G.: grinblat@df.uba.ar, email for A.V.B.: bragas@df.uba.ar.

Content:

- S1. Refractive index and extinction coefficient for crystalline GaP.
- S2. Reference values for properties of materials used in numerical simulations.
- S3. Multipolar decomposition.
- S4. Frequency and displacement at the main mechanical resonance as a function of the disk diameter.
- S5. Frequency-domain simulation of the average displacement in the z-direction.
- S6. Propagation of acoustic waves.

## S1. Refractive index and extinction coefficient for crystalline GaP.

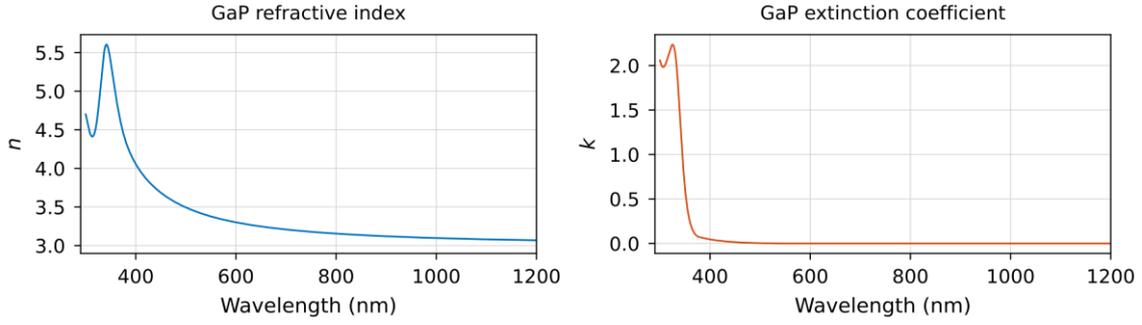

**Figure S1**. Refractive index ($n$) and extinction coefficient ($k$) of an MOCVD GaP sample obtained from ellipsometry measurements.

## S2. Reference values for properties of materials used in numerical simulations.

| Material | $n$ | $k$ | $E$ (GPa) | $v$ | $\rho$ (kg m$^{-3}$) | $\alpha$ (K$^{-1}$) | $\eta_s$ |
|---|---|---|---|---|---|---|---|
| GaP | Fig. S1 | Fig. S1 | 152.6 [1] | 0.3 [2] | 4138 [3] | 5.3×10$^{-6}$ [3] | 0.1 |
| SiO$_2$ | 1.5 | 0 | 73 [3] | 0.17 [3] | 2196 [3] | | |

**Table S1**. Reference values of properties of materials. $n$: refractive index, $k$: extinction coefficient, $E$: Young's modulus, $v$: Poisson's ratio, $\rho$: mass density, $\alpha$: coefficient of linear thermal expansion, $\eta_s$: isotropic structural loss factor.

## S3. Multipolar decomposition.

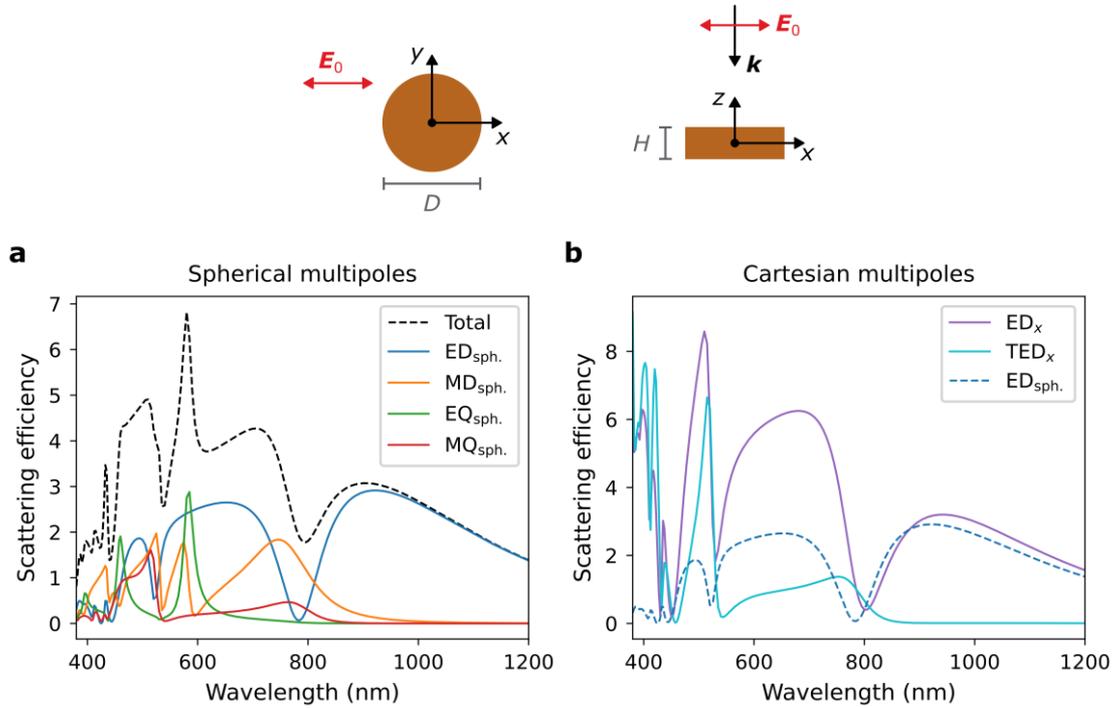

**Figure S2**. Spherical (a) and cartesian (b) multipole decompositions of the scattering efficiency of a suspended GaP nanodisk with diameter $D$ = 382 nm and height $H$ = 125 nm. ED: electric dipole, MD: magnetic dipole, EQ: electric quadrupole, MQ: magnetic quadrupole, TED: toroidal electric dipole. The subscripts *sph.* and *x* state for spherical and *x*-cartesian components, respectively. The polarization direction of the incident electric field ($E_0$) is indicated in the diagram above.

## S4. Frequency and displacement at the main mechanical resonance as a function of the disk diameter.

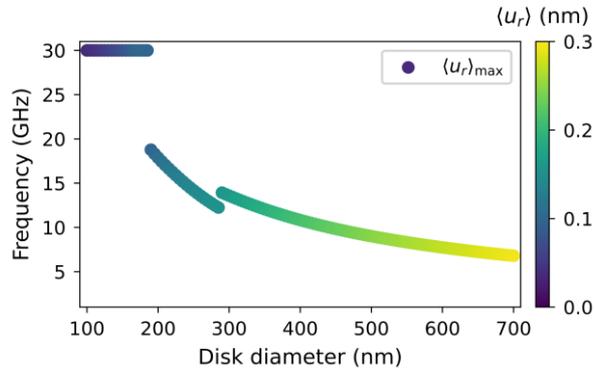

**Figure S3**. Maximum average radial displacement in the 1-30 GHz range, for different disk diameters. Same data as in Figure 4 of the main text.

## S5. Frequency-domain simulation of the average displacement in the z-direction.

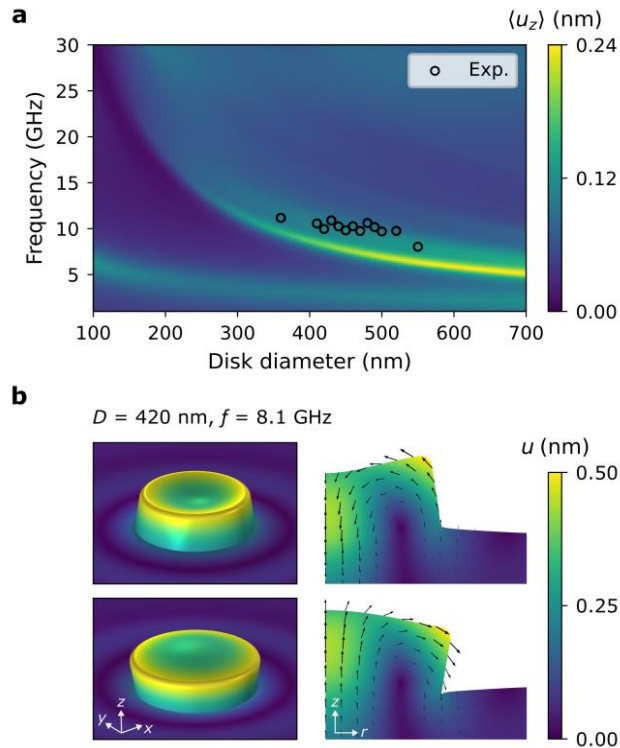

**Figure S4**. Frequency-domain simulation of the average displacement in the z-direction (see sketch in Figure 1a), $\langle u_z \rangle$, for different disk diameters. The experimental frequencies are shown as black empty circles (same data as in Figure 4a of the main text). (b) 3D view (left) and cross-section (right) of the displacement field of a 420 nm diameter nanodisk vibrating at the frequency of maximum $\langle u_z \rangle$. A scale factor 50× was applied to enlarge the deformation for visual clarity. An increase in the lattice temperature $\Delta T_L$ = 100 K was considered as the displacive excitation mechanism, and an isotropic loss factor $\eta_s$ = 0.1

was implemented to account for the internal damping. It is noted that this mode emerges when the disk is supported by a substrate and is absent in free-standing conditions.

## S6. Propagation of acoustic waves

We fitted the experimental temporal trace of the spatially decoupled pump-probe measurement (Figure S5b) using the following expression:

$$\frac{\Delta T}{T}(t) = a\left[1 + e^{-b(t-t_0)}\right]^{-1} e^{-c(t-t_0)} \sin(2\pi f(t-t_0) + p),$$

which corresponds to an exponentially damped oscillation of amplitude $a$, frequency $f$, phase $p$ and decay time $c$; modulated by a sigmoid factor with a steepness coefficient $b$, centered at the inflection point $t_0$. For more details on this empirical model, see ref. [4]. The arrival time, $t_a$, is calculated as: $t_a = t_0 - 2/b$.

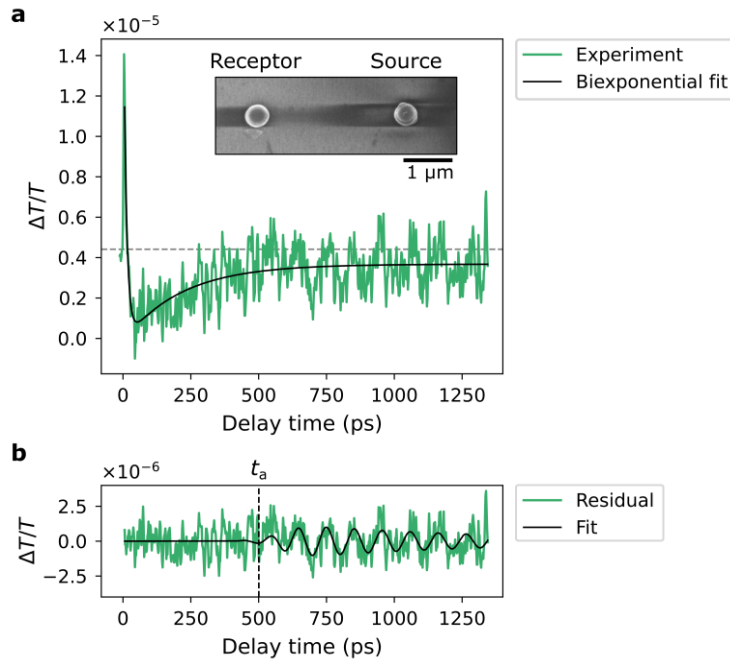

**Figure S5**. Propagating acoustic waves. (a) Spatially decoupled pump-probe measurement. The pump beam (~400 nm wavelength, 150 µW average power) illuminates a single 410 nm diameter nanodisk (source), while the probe beam (~800 nm wavelength, 15 µW average power) is temporally delayed and focused onto a nearby 420 nm diameter nanodisk (receptor). The two structures are separated by a center-to-center distance $d$ = 3 µm, as can be seen in the inset electron microscope image. The experimental signal (shown in green) corresponds to the probe differential transmission, while the black curve is a fitted biexponential decay curve. This background signal arises from a portion of the excitation spot illuminating the receiver nanostructure. (b) Residual signal after background subtraction, displaying oscillations for a probe delay time $t > t_a$, associated to the acoustic wave-driven motion of the receptor nanodisk (present only after the arrival time of the wave, $t_a$). The black line represents a curve fit by a damped sinusoid modulated by a sigmoid factor (see text). The estimated frequency and arrival time are $f$ = (9.7 ± 0.1) GHz and $t_a$ = (500 ± 60) ps, respectively. The propagation velocity was calculated as $c = d/t_a$ = (6.0 ± 0.7) nm/ps.